\documentclass[prb,twocolumn,showpacs,preprintnumbers,amsmath,amssymb,floats]{revtex4-1}
\usepackage{epsfig}
\usepackage{setspace}
\usepackage{graphicx}% Include figure files
\usepackage{amsmath}
\usepackage{epstopdf}
\usepackage{amsbsy}
\usepackage{dcolumn}% Align table columns on decimal point
\usepackage{bm}% bold math

\usepackage[usenames]{color}
\usepackage{gensymb}
\usepackage{array}

\newcommand{\vD}{\mbox{\boldmath$D$}}

\newcommand{\vL}{\mbox{\boldmath$L$}}

\newcommand{\vn}{\mbox{\boldmath$u$}}

\begin{document}

\title{Ferroelectricity in underdoped La-based cuprates}
\author{Z. Viskadourakis$^{1, 2, *}$\footnote{Corresponding author: zach@iesl.forth.gr}, S. S. Sunku$^3$, S. Mukherjee$^{4}$, B. M. Andersen$^{4}$, T. Ito$^{5}$, T. Sasagawa$^{6}$ and C. Panagopoulos$^{3,1,7, *}$}
\affiliation{
$^1$Crete Center for Quantum Complexity and Nanotechnology, University of Crete, Heraklion 71003, Greece \\
$^2$IESL-FORTH, Vassilika Vouton, Heraklion 71110, Greece\\
$^3$School of Physical and Mathematical Sciences, Division of Physics and Applied Physics, Nanyang Technological University, 637371 Singapore\\
$^4$Niels Bohr Institute, University of Copenhagen, DK-2100 Copenhagen \O, Denmark\\
$^5$National Institute of Advanced Industrial Science and Technology, Tsukuba, Ibaraki 305-8562, Japan\\
$^6$Materials and Structures Laboratory, Tokyo Institute of Technology, Kanagawa 226-8503, Japan\\
$^7$Department of Physics, University of Crete, Heraklion 71003, Greece }
%{\bf{*Corresponding authors}}: Z. V.  (zach@iesl.forth.gr) and C. P (christos@ntu.edu.sg)
%}

\begin{abstract}

We report on the observation of ferroelectricity and magnetoelectricity in La$_{2-x}$Sr$_x$CuO$_4$ and La$_2$Li$_x$Cu$_{1-x}$O$_4$. Combined with earlier works on La$_2$CuO$_{4+x}$ (Z. Viskadourakis {\it et al.}, Phys Rev. B {\bf85}, 214502 (2012)) these findings establish ferroelectricity as a new generic property in underdoped La-214 cuprates. Furthermore, the measured electric polarization can be tuned by Dzyaloshiskii-Moriya interaction resulting in a magnetoelectric effect. It is proposed that ferroelectricity results from local CuO$_6$ octahedral distortions, associated with the dopant atoms and/or clustering of the doped charge carriers, which break spatial inversion symmetry at the local scale.

\end{abstract}


\maketitle

\section{Introduction}
The phase diagram of the high temperature superconducting cuprates has been under extensive investigation since their discovery, almost three decades ago.\cite{1} In addition to high temperature superconductivity, a variety of ground states have been proposed and partly realized upon charge carrier doping. For example, the parent compound La$_2$CuO$_4$, which is an antiferromagnetic (AF) Mott insulator with N\'{e}el temperature T$_N$=325~K, \cite{2} is known to exhibit a short-range glassy phase \cite{3,4,5,6,7} and subsequently diagonal stripe order upon doping with e.g. Sr (La$_{2-x}$Sr$_x$CuO$_4$). \cite{8,9,10} Recently, research has refocused on the highly underdoped cuprates.\cite{11,12} However the evolution of the electronic ground state with the very first added charge carriers, in particular in the search for a possible broken symmetry associated with an exotic ground state and subsequent effects on the anomalous normal state properties, remains an enigma.

Although earlier efforts had in fact suggested that ferroelectricity could be present in La$_2$CuO$_{4+x}$ and YBa$_2$Cu$_3$O$_{6+x}$,\cite{13, 14} a low temperature ferroelectric (FE) phase with an associated magnetoelectric (ME) coupling was recently observed in highly underdoped La$_2$CuO$_{4+x}$ (T$_N = 320$~K).\cite{15} Notably, ferroelectricity is characterized by broken spatial inversion symmetry and typically emerges in the absence of mobile charge carriers. In the case of lightly oxygen doped La$_2$CuO$_{4+x}$ (T$_N$ = 320~K) it was proposed that the non-stoichiometric oxygen ions take interstitial positions in the La$_2$CuO$_4$ unit cell, causing a displacement in the apical oxygen ions of the CuO$_6$ octahedra, which are the building blocks of the La$_2$CuO$_4$ unit cell \cite{15}. Therefore, local-scale structural CuO$_6$ distortions take place, breaking the spatial inversion symmetry, resulting in the formation of local electric dipoles.\cite{16,17} These dipoles are localized around the oxygen interstitials forming charge clusters, which couple and freeze at low temperature, giving rise to an observable polar state.

Other microscopic mechanisms have also been proposed. These include a model based on polaron formation around the dopant atoms, \cite{18} a purely magnetic origin associated with Dzyaloshinskii-Moriya (DM) interaction leading to local structural distortions and concomitant local broken inversion symmetry, \cite{19} and a proposal for the emergence of electric polarization due to the formation of magnetic vortex/anti-vortex pairs at the edges of oriented stripe segments. \cite{20} The presence of the ME effect in La$_2$CuO$_{4+x}$ \cite{15} has also been studied by Landau theory, which includes a bi-quadratic coupling between the electric polarization and the magnetic order.\cite{21, 22} At present, there is no consensus on the origin of the FE phase in these materials. The question arising therefore is whether ferroelectricity is present in La$_2$CuO$_4$ when different dopant ions are introduced into the lattice. To shed further light on the nature of ferroelectricity and magnetoelectricity in La$_2$CuO$_4$ system, it is important to study other members of the La-214 cuprate family, especially the case where charge carriers originate from Sr and Li doping. Both of these ions occupy stoichiometric positions in the La$_2$CuO$_4$ unit cell and therefore, no dipole moments are directly associated with the dopant sites, in contrast to the case of interstitial excess oxygen ions in La$_2$CuO$_{4+x}$. \cite{15}

Here, we report measurements of the electric polarization on Sr and Li doped La$_2$CuO$_4$ single crystals. We show that La$_{1.999}$Sr$_{0.001}$CuO$_4$ exhibits distinct FE behavior along different crystallographic directions. Similar behavior is also shown in lightly oxygen doped La$_2$CuO$_{4+x}$ samples (different samples than those investigated in Ref.~\onlinecite{15}) studied here for direct comparison to La$_{1.999}$Sr$_{0.001}$CuO$_4$. The magnetic field dependence of the electric polarization is anisotropic and it can be tuned through the DM interaction. In La$_2$Li$_x$Cu$_{1-x}$O$_4$ ($x=0.01$ and $x=0.04$), the measured electric polarization is in the order of $\mu$C cm$^{-2}$ i.e., several times higher than for La$_{1.999}$Sr$_{0.001}$CuO$_4$ (as shown below) and La$_2$CuO$_{4+x}$.\cite{15} The electric polarization in La$_2$Li$_x$Cu$_{1-x}$O$_4$ is also tunable by an applied magnetic field similar to La$_{1.999}$Sr$_{0.001}$CuO$_4$ and La$_2$CuO$_{4+x}$. These results taken together demonstrate that the FE phase is present in all underdoped La-214 cuprates investigated so far, and is prompted by charge carrier doping. We propose ferroelectricity may originate from a mechanism that breaks inversion symmetry by local structural distortions of the CuO$_6$ octahedra due to the presence of the dopant ions and/or clustering of the added holes. Such local symmetry breaking induces electric dipoles which couple and order at low temperature.

\section{Methods}

The experiments were performed on twinned single crystals. La$_{1.999}$Sr$_{0.001}$CuO$_4$ crystals were grown using the laser-diode heated floating zone method \cite{23}. High purity La$_2$O$_3$ (99.99 \%), SrCO$_3$ (99.9 \%) and CuO (99.9 \%) were used as starting materials. Sr ions take La stoichiometric positions and thus the Sr concentration (x = 0.001) is higher than the purity of La$_2$CO$_3$. On the other hand, the Sr concentration is comparable to the purity of CuO. Furthermore, the crystals were carefully annealed at 1000\degree~C for 20 h under 1ppm O$_2$ - Ar flowing atmosphere. The crystal axes were determined by x-ray Laue backscattering technique. The sharp antiferromagnetic transition at 312~K (see supplementary S1) is suggestive of homogeneous distribution of the dopant ions (either Sr or excess oxygen ions that might be included in the unit cell) into the La$_{1.999}$Sr$_{0.001}$CuO$_4$ samples as any inhomogeneous distribution of the ion dopants would cause either the suppression - broadening of the AF peak at 312~K, or the appearance of secondary magnetic peaks at temperatures lower than 312~K.
\begin{figure*}[t]
\begin{minipage}{1.4\columnwidth}
\includegraphics[clip=true,width=0.99\columnwidth]{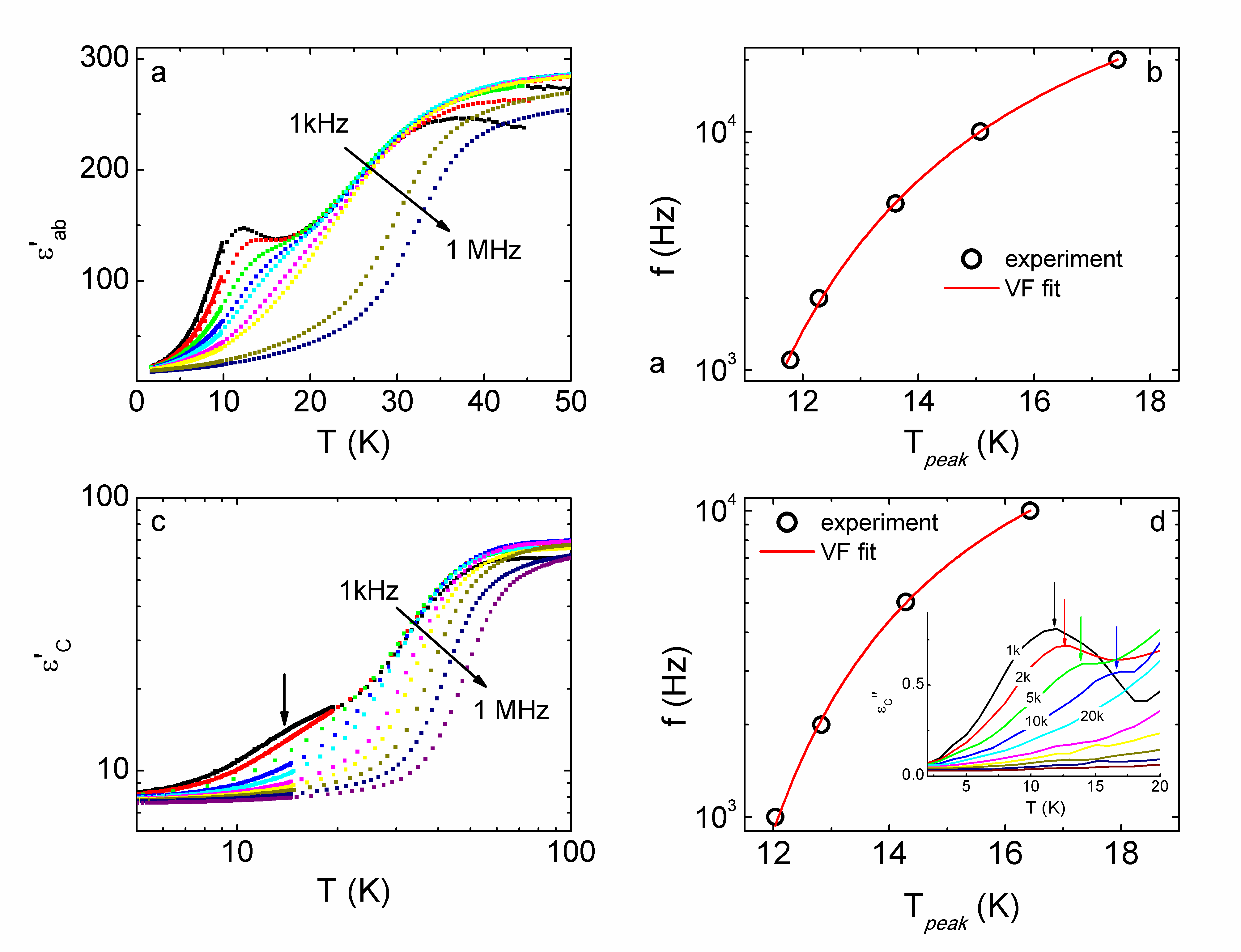}
\end{minipage}
\caption{(Color online) a. In-plane dielectric permittivity $\epsilon'_{ab}$ as a function of temperature for various frequencies for La$_{1.999}$Sr$_{0.001}$CuO$_{4+y}$. b. Experimental data for frequency f vs. peak temperature T$_{peak}$ as extracted from panel (a) (black open circles). The red solid line corresponds to the Vogel-Fulcher fit (see main text). c. Out-of-plane dielectric permittivity $\epsilon'_c$ as a function of temperature for various frequencies. The black arrow indicates the position of the low temperature dielectric peak, for 1kHz.  d. The experimental data (black open circles) for the f - T$_{peak}$ curve, are extracted from the imaginary part of the dielectric permittivity $\epsilon''_{ab}$ (inset), while the red solid line depicts the Vogel-Fulcher fit.} \label{fig:1}
\end{figure*}

To overcome the difficulty in quantifying experimentally the very low Sr concentration, we synthesized (crystal growth and annealing) oxygen doped La$_2$CuO$_{4+x}$ using the same method as La$_{1.999}$Sr$_{0.001}$CuO$_4$. \cite{23} The excellent agreement in the magnetic transition (supplementary S1), indicate similar excess oxygen doping in both materials, as the very small Sr doping will not affect the magnetic transition temperature of La$_{1.999}$Sr$_{0.001}$CuO$_4$ significantly \cite{24, 25} (therefore La$_{1.999}$Sr$_{0.001}$CuO$_{4+y}$ is the appropriate chemical formula for the Sr-doped La$_2$CuO$_4$ samples). Any possible difference in their transport properties therefore would be attributable to Sr doping.

La$_2$Cu$_{1-x}$Li$_x$O$_4$ single crystals were grown using the conventional Lamp Heated Floating Zone Technique. \cite{26} High purity La$_2$O$_3$ (99.999 \%), Li$_2$CO$_3$ (99.99 \%) and CuO (99.99 \%) were used as starting materials in order to achieve the lowest possible impurity level. The crystal axes were again determined by x-ray Laue backscattering technique. The samples were annealed at 900\degree~C for 48h, in flowing Ar atmosphere. The Li concentration was estimated from x-ray diffraction measurements of the lattice constants. \cite{26}
All crystals were cut appropriately and pairs of plate-like samples were extracted. Each pair consists of a sample with the thinnest direction along the c-axis and another with the thinnest direction along the ab-plane. The magnetization of the samples was measured using a commercial MPMS Quantum Design SQUID magnetometer, in Heraklion. The La$_{1.999}$Sr$_{0.001}$CuO$_{4+y}$ crystals exhibit T$_N$ = 312 K, whereas for the La$_2$CuO$_{4+x}$ T$_N$ = 313~K. For La$_2$Li$_x$Cu$_{1-x}$O$_4$ crystals, T$_N$ = 255~K ($x$ = 0.01) and T$_N$ = 150~K ($x$ = 0.04), respectively. The impedance and loss of the samples were measured using an LCR meter, over a frequency range 21 Hz - 2 MHz, and the dielectric permittivity was extracted, as described elsewhere.\cite{6} Electric polarization measurements were performed using two different home-built experimental stations in Heraklion and Singapore employing the pyrocurrent technique. Details of the technique, the measurement protocol and the pyrocurrent data processing can be found in the supplementary (sections S2 - S4).

\begin{figure}[]
\begin{minipage}{.99\columnwidth}
 \includegraphics[clip=true,width=0.99\columnwidth]{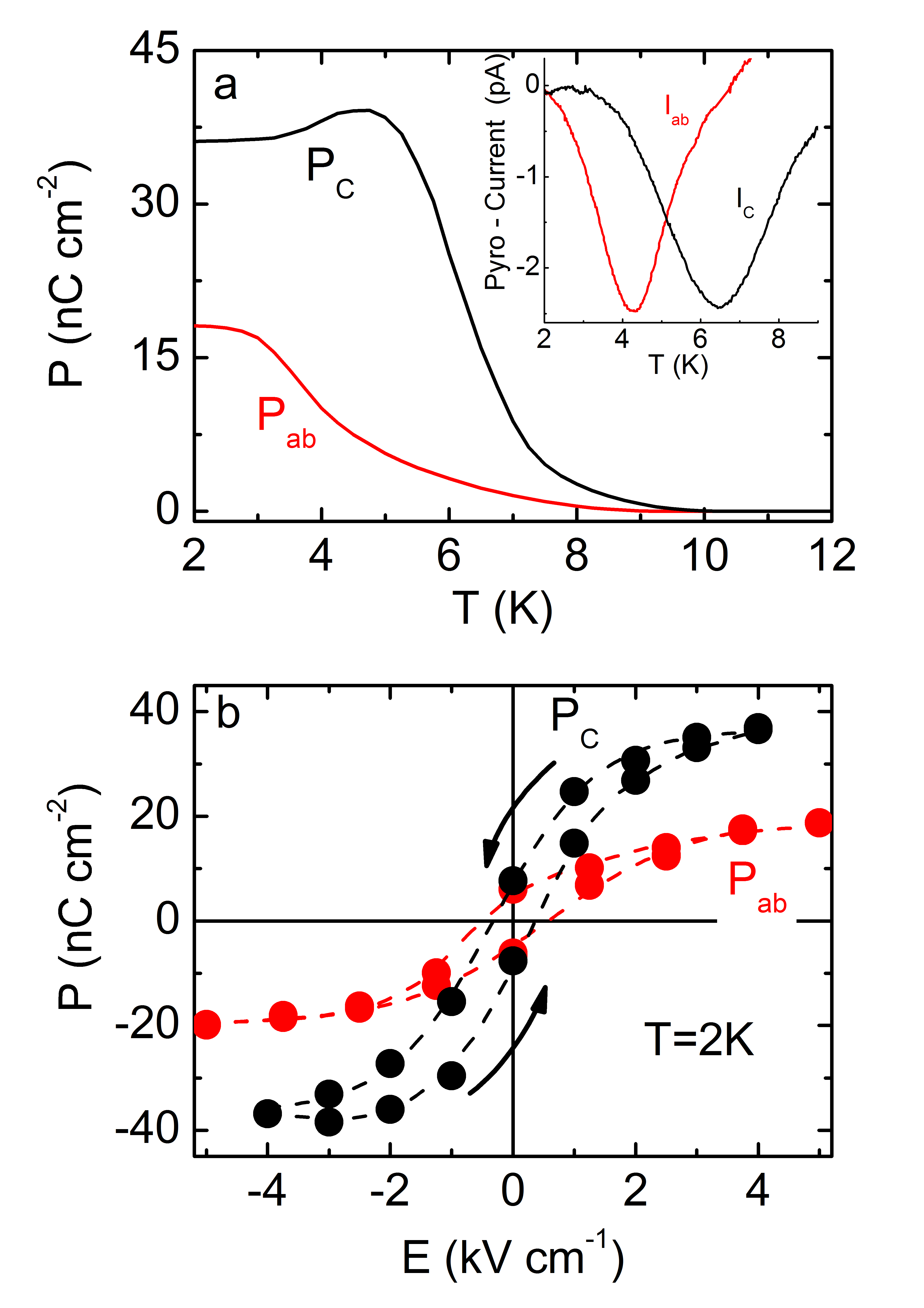}
\end{minipage}
\caption{(Color online) a. In-plane P$_{ab}$ (red line) and out-of-plane P$_c$ (black line) electric polarization as a function of temperature for La$_{1.999}$Sr$_{0.001}$CuO$_{4+y}$. Inset. The corresponding pyroelectric current curves are shown. b. Polarization as a function of applied electric field (P-E) hysteresis loops for both P$_{ab}$ (red solid circles) and P$_c$ (black solid circles) at 2~K. Broken lines are a guide to the eye.} \label{fig:2}
\end{figure}

\section{Results and Discussion}

Figure~\ref{fig:1}a shows the real part of the in-plane dielectric permittivity $\epsilon'_{ab}$ for La$_{0.999}$Sr$_{0.001}$CuO$_{4+y}$. At high temperatures, a step-like decrement is observed for all measured frequencies. This decrement shifts to higher temperature with increasing frequency f, indicative of a common dielectric relaxation process. At low frequencies, an additional dielectric peak develops which shifts to higher temperature and is suppressed with increasing f. Such a peak in the dielectric permittivity can be attributed either to an intrinsic charge relaxor characterized by a diffused phase transition or to spurious effects arising from the electrical contacts such as extrinsic Maxwell-Wagner phenomena. \cite{27}

To reveal the intrinsic character of the dielectric peaks, measurements of the dielectric permittivity were repeated several times with renewed contacts, giving consistent experimental results in the temperature regime around the peaks. Thus the electrical contacts do not affect the measured dielectric permittivity. Also, both the temperature T$_{peak}$ where the dielectric peak occurs and the magnitude of the permittivity at the peak position $\epsilon'_{peak}$ cannot be described by the empirical relations reported for pseudo-FE relaxor behavior. \cite{15, 27} Nevertheless, such a relaxation process due to the slowing of polar clusters, can be described by the Vogel-Fulcher (VF) relation $f = f_o \exp [ -E_a / k_B ( T - T_{fr} ) ]$, where the characteristic freezing temperature T$_{fr}$ corresponds to the estimated temperature below which the polar clusters freeze. \cite{28} Notably, electric polarization may emerge below T$_{fr}$ in relaxor ferroelectrics. \cite{29} Figure 1b shows f as a function of T$_{peak}$ for La$_{1.999}$Sr$_{0.001}$CuO$_{4+y}$. We estimated the T$_{fr-ab}$ = (7.3 $\pm$ 0.3) K by fitting the experimental data against the VF relation. Similar analysis of the imaginary part \cite{30} of the out-of-plane dielectric permittivity $\epsilon''_c$ gives a freezing temperature T$_{fr-c}$ = (8.6 $\pm$ 0.5) K, as (Fig~\ref{fig:1}c and Fig~\ref{fig:1}d). Therefore, the observed behavior could be due to a FE relaxor behavior, characterized by a diffused phase transition and the freezing of short-range cluster-like order. The above data analysis also reveals an anisotropic behavior in the charge dynamics in agreement with an earlier report.\cite{6}

Figure~\ref{fig:2} shows the in-plane (P$_{ab}$) and the out-of-plane (P$_c$) electric polarization as a function of temperature for La$_{1.999}$Sr$_{0.001}$CuO$_{4+y}$. For both directions, the polarization increases with decreasing temperature, below 10 K. P$_c$ and P$_{ab}$ exhibit distinct temperature dependencies. Furthermore, corresponding pyroelectric curves (Fig~\ref{fig:2}a inset) show that the pyroelectric current local minima occur at different temperatures, namely 4 K and 6.5 K, for the in-plane and the out-of-plane directions, respectively. These temperatures are also lower than the corresponding freezing temperatures extracted from VF analysis.\cite{28} Notably, a similar anisotropy has been observed in the spin-glass temperature. \cite{9, 31}  Furthermore, the electric polarization curves reverse with reversing the polarity of the applied electric field.

In proper FE's the onset of ferroelectricity is defined as the temperature above which there is no measurable polarization and the corresponding pyroelectric current exhibits a sharp minimum. On the other hand, in relaxor FEs a transition occurs over a large time scale. Similar to our data, the pyroelectric minimum is not sharp, resulting in a broad transition in the polarization with respect to temperature, therefore preventing a precise assignment of a transition temperature. We assign the transition towards a FE state as the temperature where a minimum (albeit broad) occurs in the pyroelectric current. This assignment is corroborated by the fact that the temperature where the local minimum occurs is robust against both the electric field applied during cooling and the temperature sweep rate.

The P-E hysteresis loops for both P$_c$ and P$_{ab}$, are shown in Fig~\ref{fig:2}b (T = 2 K). The raw data yield a sizable anisotropic electric polarization, with the in-plane and out-of-plane values reaching P$_{ab}$ = 18 nCcm$^{-2}$ and P$_c$ = 36 nCcm$^{-2}$ respectively. The dielectric anomaly and the relaxor behavior in the charge dynamics indicate a dominant electronic contribution towards the polarization. In addition, the observation of non-zero polarizations along both the c-axis and ab-plane implies that the electric polarization may arise from an ensemble of charge clusters with varying polarization vectors.

\begin{figure}[]
\begin{minipage}{.99\columnwidth}
\includegraphics[clip=true,width=0.99\columnwidth]{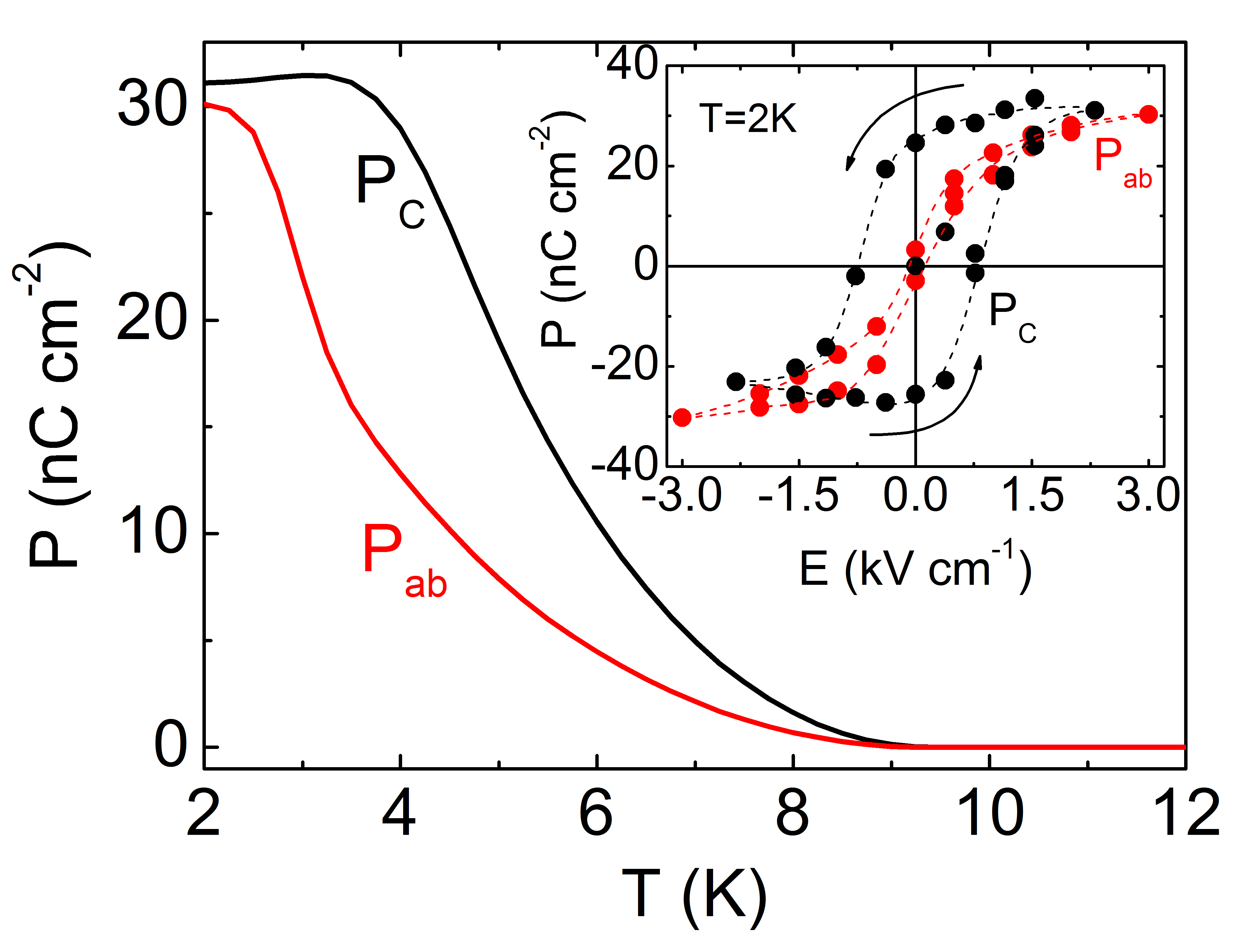}
\end{minipage}
\caption{(Color online) In-plane P$_{ab}$ (red line) and out-of-plane P$_C$ (black line) electric polarization as a function of temperature for lightly oxygen-doped La$_2$CuO$_{4+x}$ (T$_N$ = 313K). Inset. Polarization as a function of applied electric field (P-E) hysteresis loops for P$_{ab}$ (red symbols) and P$_c$ (black symbols) at 2~K. The dash lines are a guide to the eye.  } \label{fig:3}
\end{figure}

Although the intrinsic nature of the different trends in our observations for La$_{1.999}$Sr$_{0.001}$CuO$_{4+y}$ (T$_N$ = 312 K, n $\sim$ 10$^{18}$ cm$^{-3}$),\cite{32} it is not clear whether they come from Sr doping or from the excess oxygen ions included in the La$_2$CuO$_4$ unit cell (as discussed before). To this point, we performed electric polarization measurements on lightly oxygen-doped La$_2$CuO$_{4+x}$ with T$_N$ = 313 K (n $\sim$ 10$^{18}$ cm$^{-3}$).\cite{32} Both P$_{ab}$ and P$_c$ increase (Fig~\ref{fig:3}) with decreasing temperature, however P$_c$(T) exhibits a different temperature dependence compared to P$_{ab}$(T). Namely, a change in slope is observed at 5 K and 3 K, for out-of-plane and in-plane directions respectively, coinciding with the corresponding pyrocurrent local minima. These values are also lower than the charge freezing temperatures obtained from the VF analysis (supplementary S5). At T = 2 K, P$_c$ is almost equal to P$_{ab}$. Though, both the remnant electric polarization (P (E = 0)) and the coercive electric field (E (P=0)) differ for the two measured orientations (inset Fig~\ref{fig:3}).  Electric polarization appears stronger in the out-of-plane direction (i.e. larger remnant electric polarization and coercive electric field compared to the ab-plane). We note that a finite conductivity could lead to an expanded "banana-like" P-E loop.\cite{33} However, this is not the case in the P-E loops depicted in the inset of fig. 3. Electrical transport experiments in La$_2$CuO$_4$ single crystals \cite{34} reveal that the in-plane conductivity is considerably larger than the out-of-plane counterpart. Such high conductivity could result in a relatively lossy dielectric, which can manifest itself as a "chubby" banana-like P-E curve, in the ab-direction. However, in our case the in-plane P-E loop is slimmer than the out-of plane loop, indicating the intrinsic nature of both the enhanced out-of-plane remnant electric polarization and the coercive electric field.

The above experimental evidence indicates that Sr doping contributes to the measured electric polarization. The distinct dielectric permittivity, anisotropy, remnant polarization and coercive field between La$_2$CuO$_{4+x}$ and La$_{1.999}$Sr$_{0.001}$CuO$_{4+y}$ indicate distinguishing effects between oxygen and Sr doping. We also measured La$_{1.998}$Sr$_{0.002}$CuO$_{4+y}$ however the material is not sufficiently insulating to permit transport measurements of the electric polarization.  Hence, although possible excess oxygen may still affect the observations in La$_{1.999}$Sr$_{0.001}$CuO$_{4+y}$, the above mentioned observations indicate a characteristic effect due to Sr doping (supplementary table S1).

We now turn to the effect of the applied magnetic field. The spins in the AF phase of La-214 are weakly canted out-of-plane due to the presence of a finite DM interaction.\cite{35} Spin canting causes a weak ferromagnetic (WF) moment along the c-axis in each CuO$_2$ plane, although the opposite spin canting in alternate planes leads to a net cancelation of the WF moments. By applying an external magnetic field along the c-axis, the WF moments can be aligned in the same direction above a critical magnetic field H$_{cr}$, which for our La$_{1.999}$Sr$_{0.001}$CuO$_{4+y}$ samples is 6~T (supplementary section S6).

\begin{figure}[]
\begin{minipage}{.99\columnwidth}
\includegraphics[clip=true,width=0.99\columnwidth]{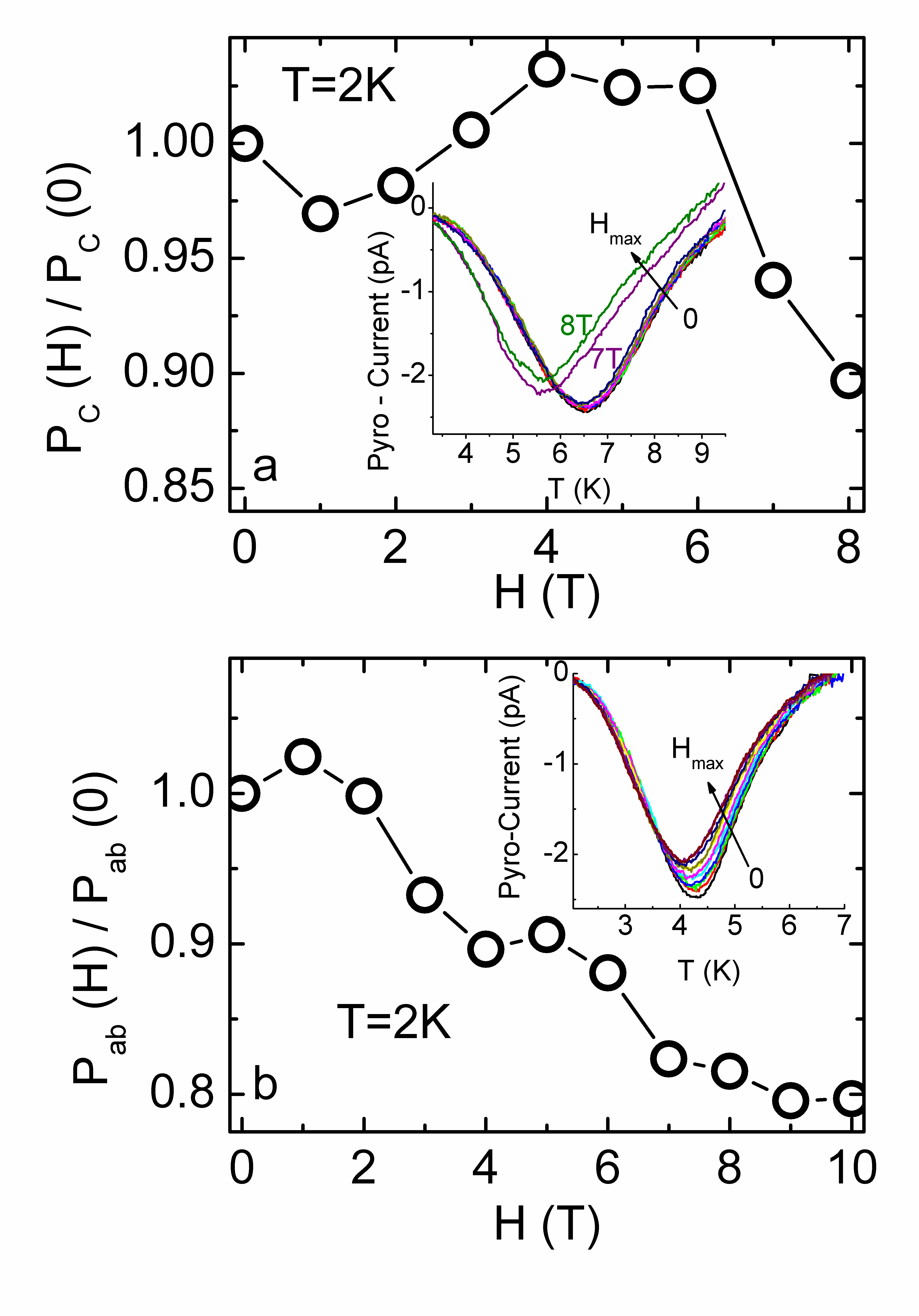}
\end{minipage}
\caption{(Color online) a. Out-of-plane polarization P$_c$ as a function of applied magnetic field H at T=2~K for La$_{1.999}$Sr$_{0.001}$CuO$_{4+y}$ (E $\|$c and H $\|$ c). Inset. Pyroelectric current as a function of temperature for various H. b. P$_{ab}$ as a function of H (E $\|$ ab, H $\|$ ab). Inset. Pyroelectric current as a function of temperature for various H.
} \label{fig:4}
\end{figure}

Figure~\ref{fig:4}a shows P$_c$ as a function of applied magnetic field (here E$\|$c and H$\|$c). P$_c$ decreases abruptly above H$_{cr}$ = 6 T and the pyroelectric current minimum (inset in Fig~\ref{fig:4}a) shifts to lower temperatures, indicating a suppression of the FE state as the material enters the WF state (such abrupt changes are also observed for H$\|$c and E$\perp$c). On the other hand, both P$_{ab}$ and the pyrocurrent minimum (E$\|$ab and H$\|$ab) decrease smoothly with increasing H, as seen in Fig~\ref{fig:4}b (similar changes are observed for E$\perp$ab and H$\|$ab). Hence, the FE order in La$_{1.999}$Sr$_{0.001}$CuO$_{4+y}$ is coupled to the underlying AF structure and is influenced by the DM interaction in a manner similar to the recent observations for La$_2$CuO$_{4+x}$. \cite{15, 20, 21}

The similarity in the FE state among La$_2$CuO$_{4+x}$ and La$_{1.999}$Sr$_{0.001}$CuO$_{4+y}$ cuprates is notable even though oxygen and Sr doping do not affect the CuO$_6$ structure in the same way; while the excess oxygen take non-stoichiometric positions, Sr ions substitute for La. Because Sr and oxygen dopants take positions outside the oxygen octahedra it is important to study the evolution of ferroelectricity in the La-214 family of cuprates using also Li doping, which directly replaces Cu ions in the CuO$_6$ octahedra. Among other dopants, such as Mg and Zn which can be used to replace Cu, Li ions exhibit the largest ionic radius (0.76 \AA) compared to 0.74 \AA~for Zn, 0.72 \AA~for Mg and 0.73 \AA~for Cu. Furthermore, La$_2$Li$_x$Cu$_{1-x}$O$_4$ remains insulating up to 4 \%, \cite{26, 36} allowing pyrocurrent measurements even at relatively high dopings.

\begin{figure}[]
\begin{minipage}{.99\columnwidth}
\includegraphics[clip=true,width=0.99\columnwidth]{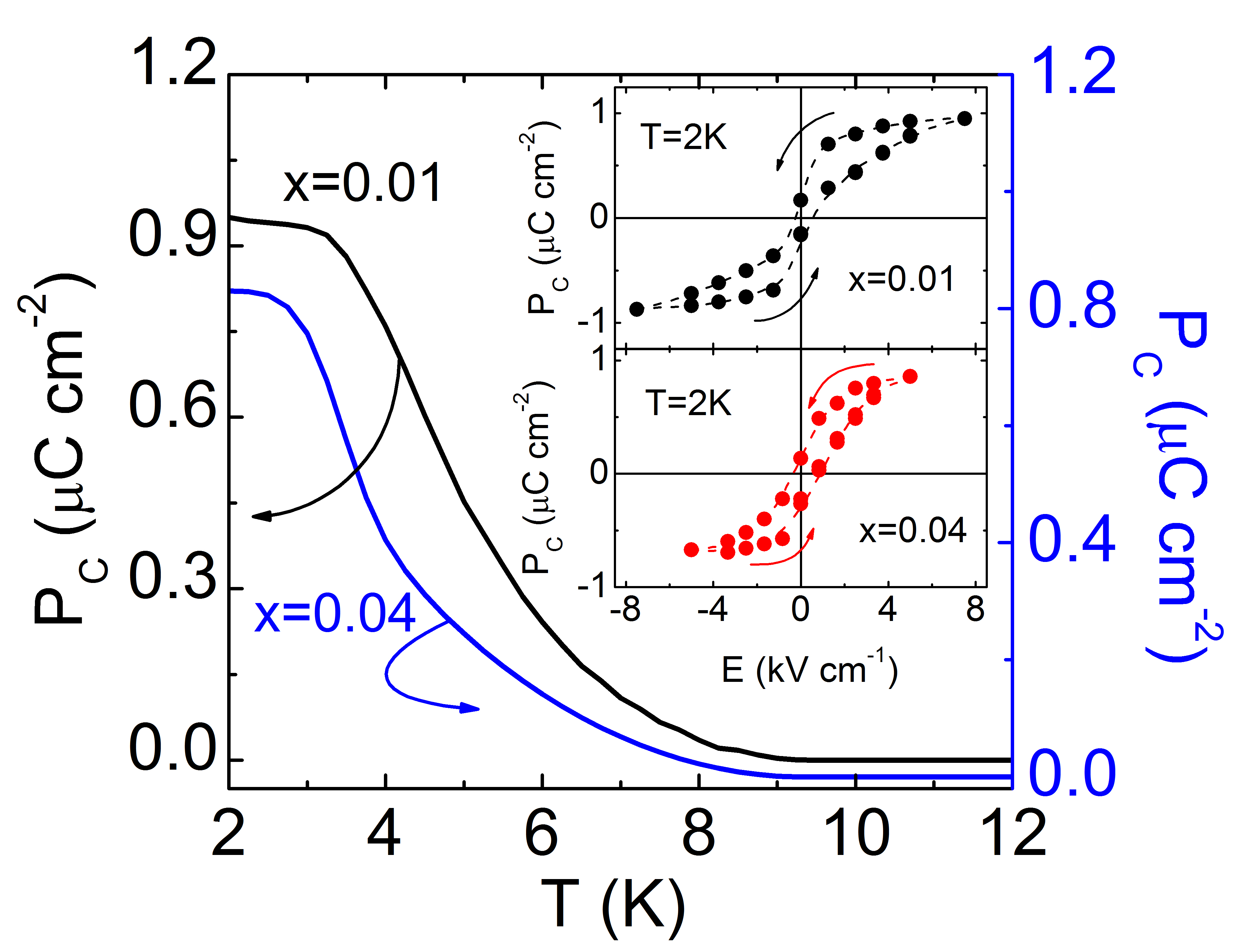}
\end{minipage}
\caption{(Color online) Out of plane polarization P$_C$ as a function of temperature for La$_2$Li$_x$Cu$_{1-x}$O$_4$ with $x$ = 0.01 (black solid line) and $x$ = 0.04 (blue solid line). Polarization as a function of applied electric field (P-E) slim loops for $x$ = 0.01 (upper inset) and $x$ = 0.04 (lower inset) at 2~K are also shown. Broken lines are a guide to the eye.} \label{fig:5}
\end{figure}

Figure~\ref{fig:5} shows the temperature dependence of P$_c$ for La$_2$Li$_x$Cu$_{1-x}$O$_4$ ($x$ = 0.01 and $x$ = 0.04). In both cases the electric polarization increases with decreasing temperature, below $\sim$9~K. A slope change in the electric polarization occurs at 5~K for $x$ = 0.01 and at 3.5~K for $x$ = 0.04, respectively. Applying the VF analysis on the corresponding dielectric permittivity data (supplementary section S7), we find a freezing temperature T$_{fr-c}$ = (5 $\pm$ 0.3) K for $x$ = 0.01. Furthermore, T$_{fr}$ is comparable to the values reported by Park {\it{et al.}} \cite{37} suggesting charge glassiness may be associated with the onset of ferroelectricity in La-214 cuprates. It was not possible to calculate the corresponding T$_{fr}$ for $x$ = 0.04, since the dielectric permittivity peaks are too broad to define a characteristic peak position (supplementary section S7). Moreover, the polarization is remarkably large namely, 900 nCcm$^{-2}$ for $x$ = 0.01 and 800 nCcm$^{-2}$ for $x$ = 0.04. Notably, there is no pyroelectric signal along the in-plane direction, in both Li-doped samples. This is probably due to the relatively high in-plane electric conductivity of the sample. \cite{6}

\begin{figure}[]
\begin{minipage}{.99\columnwidth}
\includegraphics[clip=true,width=0.99\columnwidth]{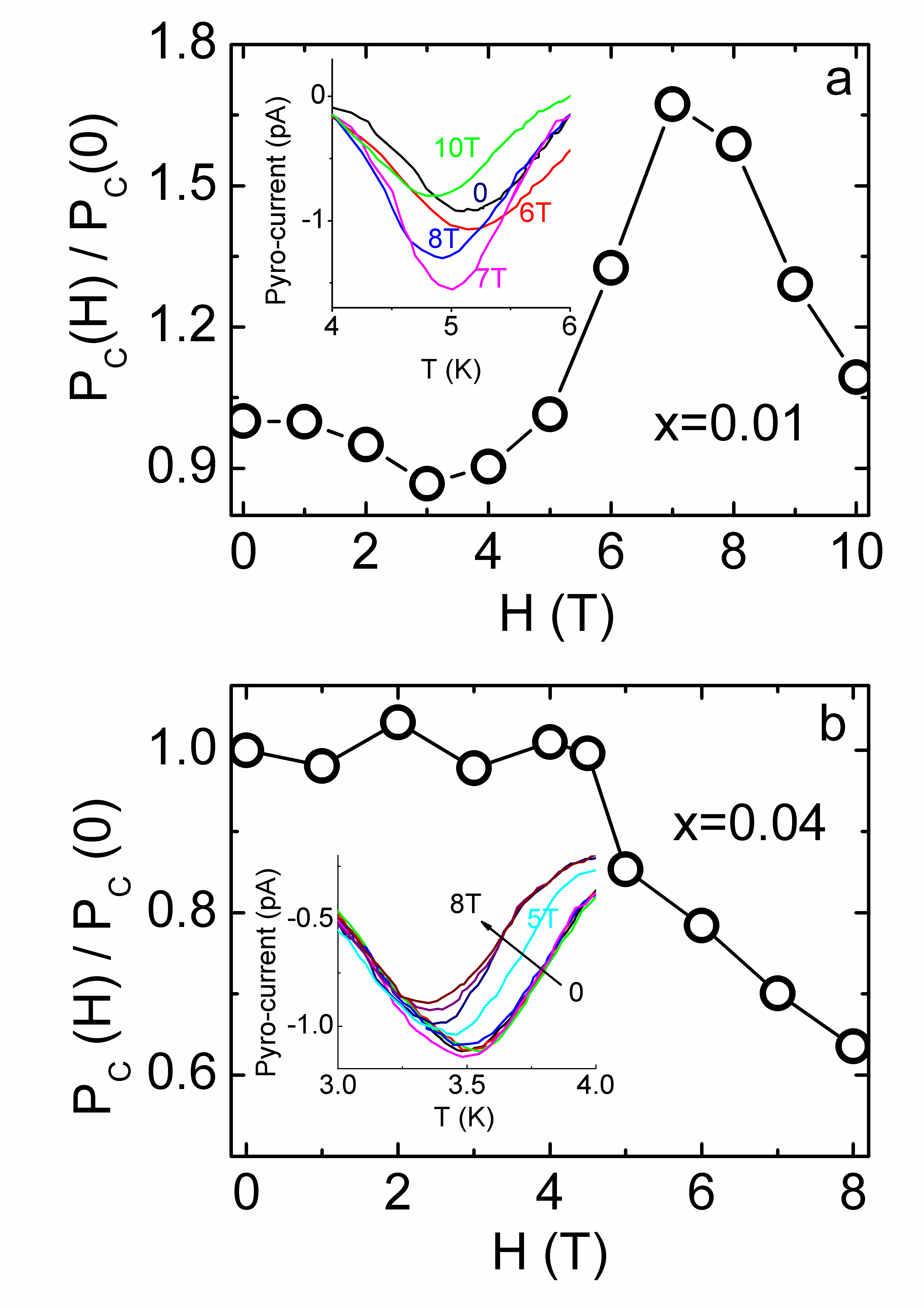}
\end{minipage}
\caption{(Color online) Out-of-plane polarization P$_C$ as a function of applied magnetic field H for a. La$_2$Li$_{0.01}$Cu$_{0.99}$O$_4$. The corresponding pyroelectric current is shown in the inset of the graph, for selected magnetic fields. b. P$_C$ vs. magnetic field for La$_2$Li$_{0.04}$Cu$_{0.96}$O$_4$. The corresponding pyroelectric signal is shown in the inset. In both cases E $\|$ c and H $\|$ c.} \label{fig:6}
\end{figure}

Figure~\ref{fig:6} shows the magnetic field dependence of P$_c$ for both La$_2$Li$_x$Cu$_{1-x}$O$_4$ samples at T = 2~K (E $\|$ c and H $\|$ c). For $x$ = 0.01, P$_c$ is enhanced below H$_{cr}$ = 6.5 T, increasing by almost 1.7 times its zero-field value (Fig~\ref{fig:6}a) before getting suppressed. Theoretical studies suggest this effect may arise due to a DM induced magnetoelectric coupling. \cite{20, 21} For $x$ = 0.04, (H$_{cr}$ = 4 T) P$_c$ is roughly constant up to H$_{cr}$, while it decreases when the WF state sets in. Notably, the FE transition temperature is also suppressed above the WF transition for both doping levels (inset Fig~\ref{fig:6}a and Fig~\ref{fig:6}b), similar to La$_{1.999}$Sr$_{0.001}$CuO$_{4+y}$. Therefore, La$_2$Li$_x$Cu$_{1-x}$O$_4$ behaves qualitatively similar to La$_{1.999}$Sr$_{0.001}$CuO$_{4+y}$ and La$_2$CuO$_{4+x}$ indicating ferroelectricity is a common ground state in the La-214 cuprates with an associated magnetoelectricity tunable by DM interaction.

According to the scenario proposed earlier, in La$_2$CuO$_{4+x}$ charge clusters give rise to ferroelectricity. \cite{15, 20, 21} Although we cannot rule out alternative mechanisms for the emergence of ferroelectricity, our proposal is likely to apply also to La$_{1.999}$Sr$_{0.001}$CuO$_{4+y}$ and La$_2$Li$_x$Cu$_{1-x}$O$_4$, since local-scale CuO$_6$ octahedral distortions may cause the spatial inversion symmetry breaking resulting in the formation of electric dipoles. These dipoles are in turn localized around the dopant ions creating polar clusters, which fluctuate and with decreasing temperature freeze resulting in a measurable ferroelectricity. Theoretical and experimental evidence for both Sr and oxygen-doped La$_2$CuO$_4$ also indicate the presence of local octahedral distortions correlating to charge inhomogeneities. \cite{38,39,40,41,42} Cordero {\it{et al.}} reported the development of a significant tilt mode associated with oxygen octahedral distortion below T = 10~K from anelastic spectroscopy measurements on nearly undoped La$_{2-x}$Sr$_x$CuO$_4$ crystals. \cite{17} It was argued that these tilt modes are due to fluctuating low temperature tetragonal structures at domain walls of the low temperature orthorhombic lattice, and are coupled to hole clusters. Additionally, a non-centrosymmetric monoclinic distortion in the CuO$_6$ octahedra in orthorhombic La$_{2-x}$Sr$_x$CuO$_4$ and La$_2$CuO$_{4+x}$ has also been observed. \cite{16} It is therefore plausible that the observed ferroelectricity may be linked to local-scale distortions from the ideal octahedral structure.  We note however, further investigation is necessary to narrow down the possibilities for the physical mechanism driving our observations. \cite{15,18,19,20,21,22}

\section{Summary}

We report evidence for the presence of FE and ME coupling in underdoped La-214 cuprates, namely La$_2$CuO$_{4+x}$, La$_{1.999}$Sr$_{0.001}$CuO$_{4+y}$ and La$_2$Li$_x$Cu$_{1-x}$O$_4$ ($x$ = 0.01 and $x$ = 0.04). For both Sr and oxygen doped samples, we observed distinct electric polarization behavior along in-plane and out-of-plane crystallographic directions. In all cases the electric polarization is strongly influenced by DM interaction. Considering the above experimental evidence, the FE order and its related magnetoelectricity appear to be a generic property of La-214 cuprates. We propose that the observed ferroelectricity originates from the CuO$_6$ octahedral distortions, occurring around the dopant ions and/or due to the formation of charge clustering, resulting to a local-scale spatial symmetry breaking symmetry.

\section{acknowledgements}
The work in Singapore was supported by the National Research Foundation, Singapore, through a Fellowship and Grant NRF-CRP4-2008-04. The work in Greece was partially supported by the European Union's Seventh Framework Program (FP7-REGPOT-2012-2013-1) under grant agreement n316165. B.M.A. acknowledges support from the Lundbeckfond fellowship (grant A9318).

\widetext
\pagebreak

\begin{widetext}
\begin{center}
\textbf{\large Supplementary Information}\linebreak\\
Z. Viskadourakis, S. S. Sunku, S. Mukherjee, B. M. Andersen, T. Ito, T. Sasagawa and C. Panagopoulos\linebreak\\
\end{center}
\end{widetext}

\setcounter{equation}{0}
\setcounter{figure}{0}
\setcounter{table}{0}
\setcounter{page}{1}
\makeatletter
\renewcommand{\thefigure}{S\arabic{figure}}
\renewcommand{\thetable}{S\arabic{table}}
\renewcommand{\bibnumfmt}[1]{[S#1]}
\renewcommand{\citenumfont}[1]{S#1}

\begin{figure*}[]
\begin{minipage}{1.4\columnwidth}
\includegraphics[clip=true,width=0.99\columnwidth]{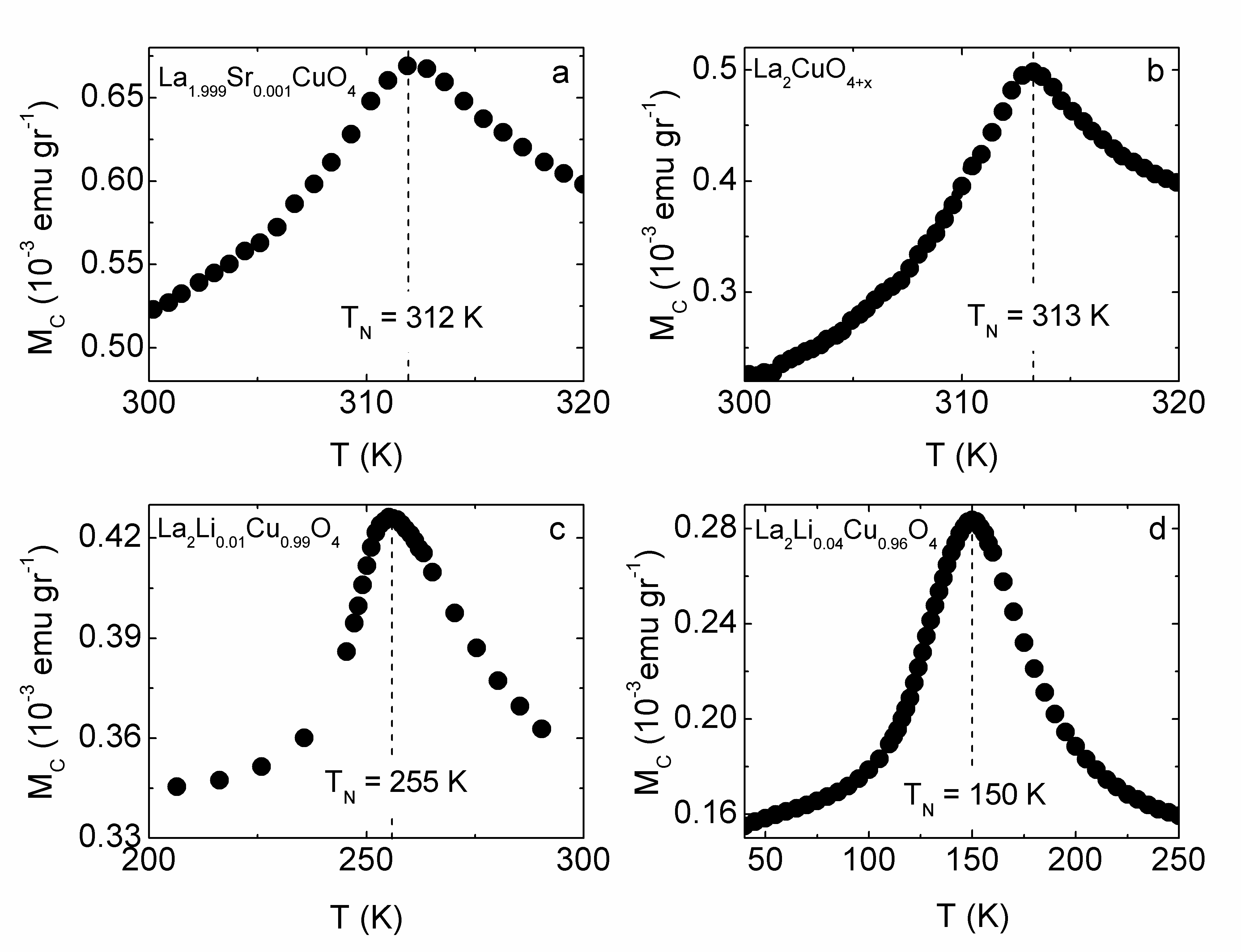}
\end{minipage}
\caption{(Color online) Bulk magnetization with the magnetic field applied along the c-axis (H = 1 kG) for a. La$_{1.999}$Sr$_{0.001}$CuO$_{4+y}$, b. lightly oxygen-doped La$_2$CuO$_{4+x}$, c.  La$_2$Li$_{0.01}$Cu$_{0.99}$O$_4$  and d.  La$_2$Li$_{0.04}$Cu$_{0.96}$O$_4$. The observed peaks correspond to the N\'{e}el transitions.} \label{fig:s1}
\end{figure*}

\begin{figure}[b]
\begin{minipage}{.99\columnwidth}
\includegraphics[clip=true,width=0.99\columnwidth]{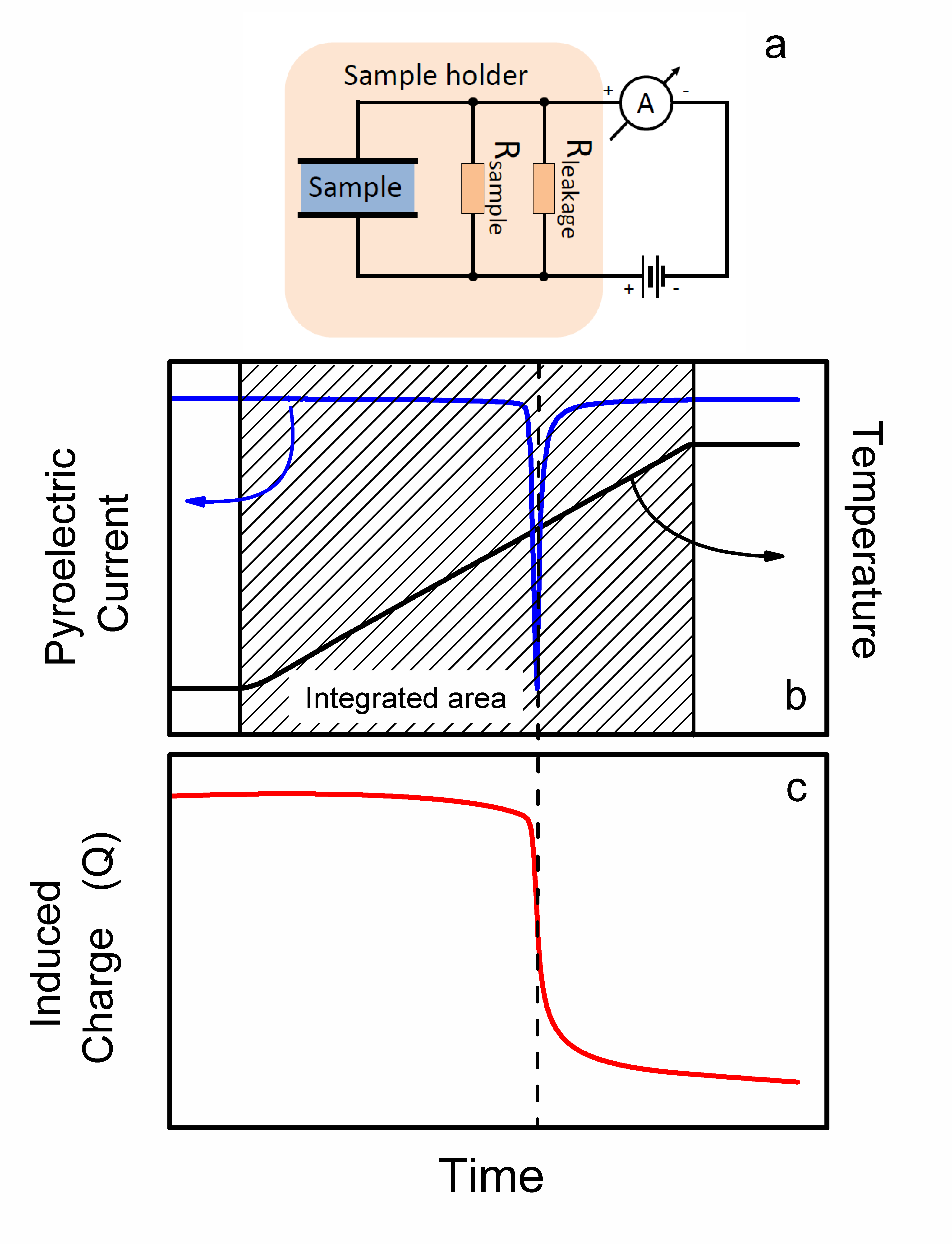}
\end{minipage}
\caption{(Color online) a. Electric circuit for polarization measurements. The sample (light blue rectangular) is placed between metallic plates. R$_{sample}$ and R$_{leakage}$ are the parasitic resistance components as described in the text. b. Pyrocurrent vs. time (blue solid line). A sharp negative peak is observed when the polarization of the sample vanishes. The temperature change is represented by the solid black line. c. The integration of the pyrocurrent peak results in the induced charge curve. Dividing by the area of the capacitor we obtain the electric polarization of the sample. In the integration process we consider the electric polarization above the ferroelectric transition temperature to be zero at the initial state.} \label{fig:s2}
\end{figure}

\begin{figure}[b]
\begin{minipage}{.99\columnwidth}
\includegraphics[clip=true,width=0.99\columnwidth]{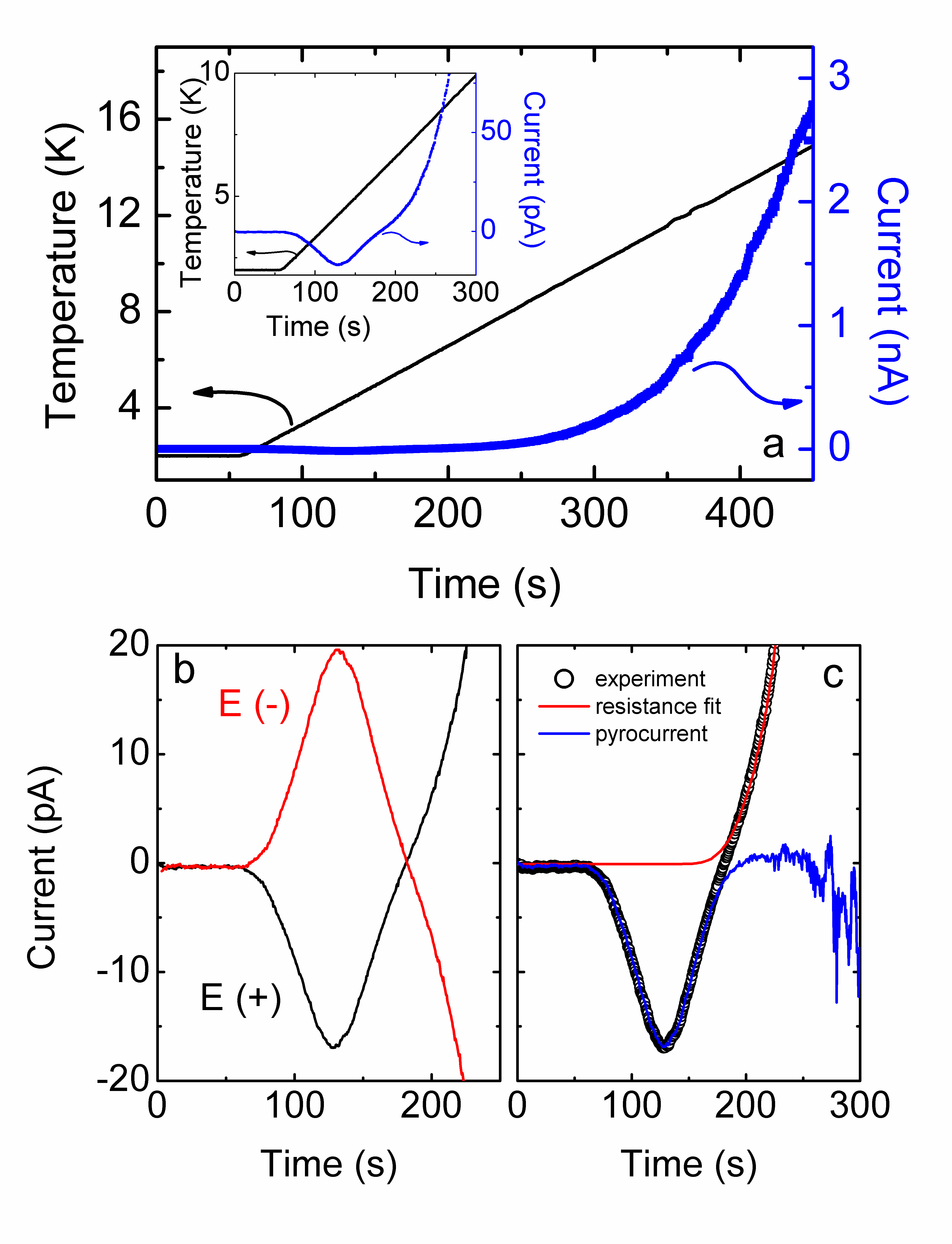}
\end{minipage}
\caption{(Color online) a. Electric current (blue solid line) and temperature (black solid line) measured during the experimental process. The inset of the graph depicts a detailed picture of the measured current. A local current minimum is recorded. b. The local minimum changes sign with respect to the electric field reversal. c. The measured electric current (black open circles) can be processed to extract the pyroelectric signal (blue solid line) to determine the electric polarization of the sample. The red solid line represents the resistive part of the measured current.} \label{fig:s3}
\end{figure}

\section*{S1. Magnetization measurements}

Figure~\ref{fig:s1}a shows the measured out-of-plane magnetization for La$_{1.999}$Sr$_{0.001}$CuO$_{4+y}$. The peak observed at T = 312 K corresponds to the N\'{e}el temperature. A similar peak is observed for the in-plane component of the magnetization again at T = 312 K. Figure~\ref{fig:s1}b shows the out-of plane magnetization for La$_2$CuO$_{4+x}$ (T$_N$ = 313 K). Figure~\ref{fig:s1}c and Fig~\ref{fig:s1}d show the out-of plane magnetization for the Li-doped samples, $x$ = 0.01 and $x$ = 0.04, respectively. The corresponding magnetization is comparable to earlier studies. \cite{s1, s2}

\begin{figure*}[]
\begin{minipage}{1.4\columnwidth}
\includegraphics[clip=true,width=0.99\columnwidth]{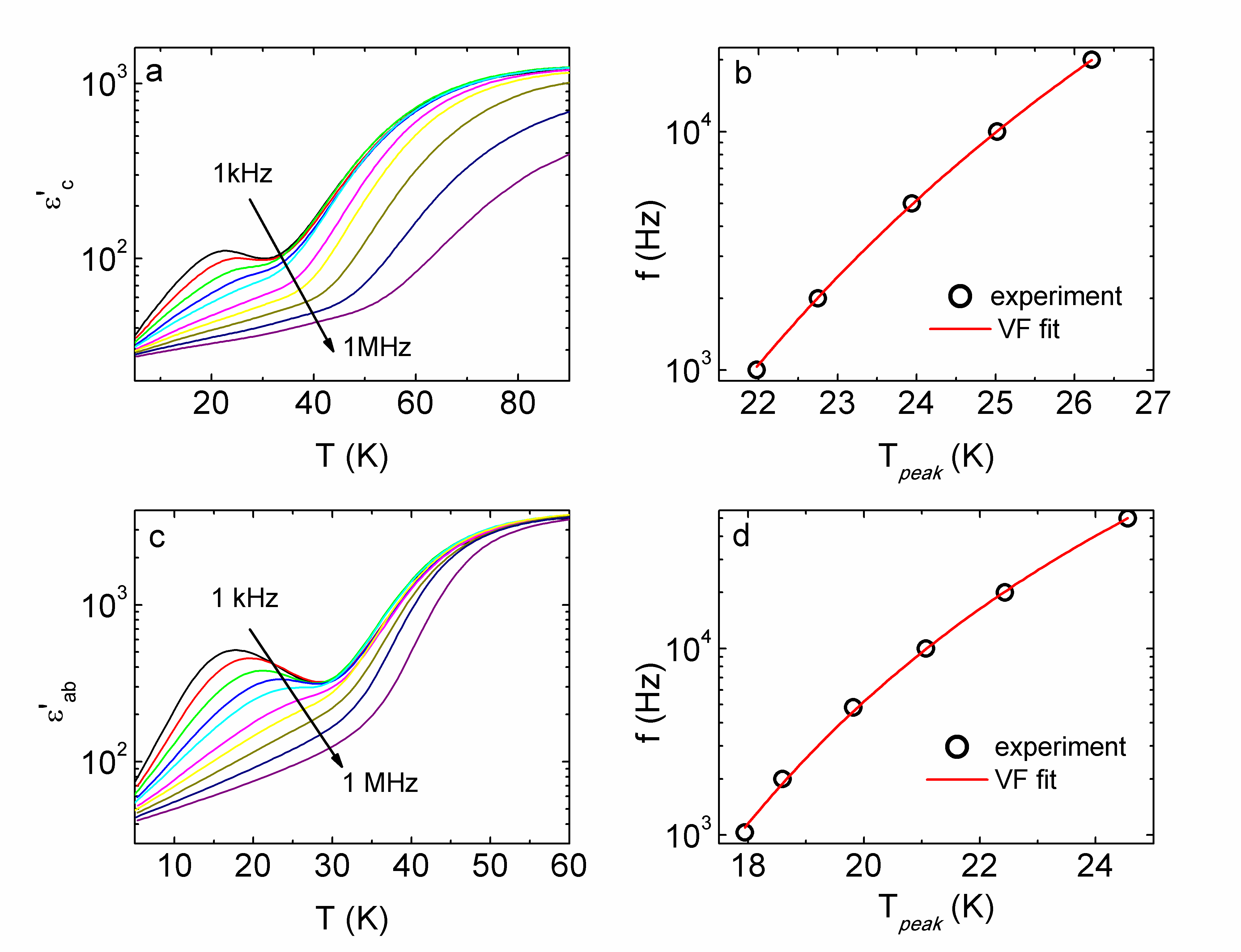}
\end{minipage}
\caption{(Color online) a. Real part of the dielectric permittivity $\epsilon'_C$ as a function of temperature, for lightly oxygen-doped La$_2$CuO$_{4+x}$ (T$_N$ = 313 K), measured for various frequencies. b. Frequency f as a function of peak temperature T$_{peak}$ extracted from panel a (black open circles). The red solid line corresponds to the Vogel-Fulcher fit. c. Real part of the in-plane dielectric permittivity $\epsilon'_{ab}$ as a function of temperature for various frequencies. d. Vogel-Fulcher fit (red solid line) applied to the f - T$_{peak}$ curve temperature data (black open circles).} \label{fig:s4}
\end{figure*}

\begin{figure*}[]
\begin{minipage}{1.4\columnwidth}
\includegraphics[clip=true,width=0.99\columnwidth]{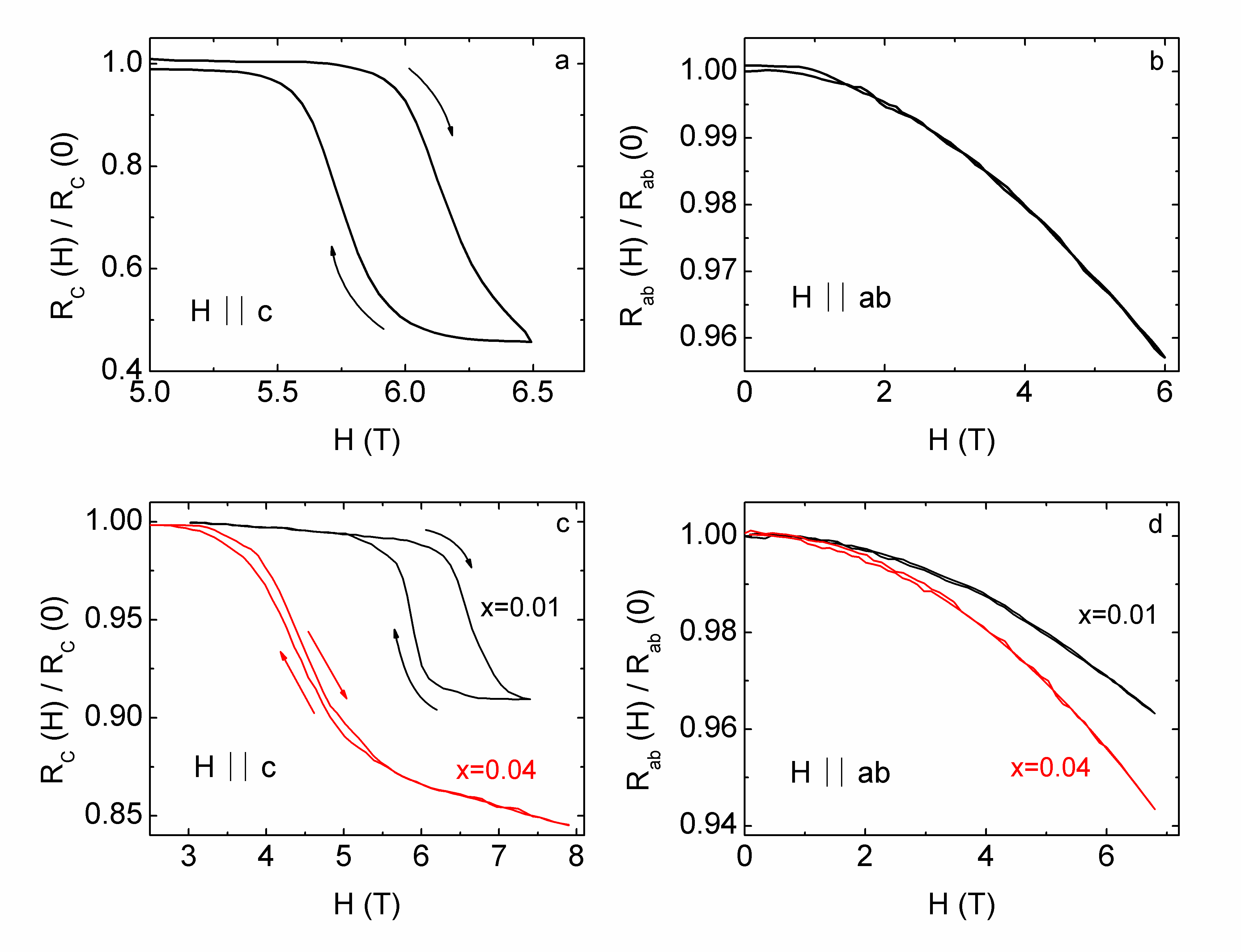}
\end{minipage}
\caption{(Color online) a. Magnetic field dependence of the out-of-plane normalized resistivity R$_C$(H) / R$_C$(0) for La$_{0.999}$Sr$_{0.001}$CuO$_{4+y}$ (T = 20 K; E$\|$c and H$\|$c). b. In-plane normalized resistivity R$_{ab}$(H) / R$_{ab}$(0) as a function of applied magnetic field (T = 20 K; E$\|$ab and H$\|$ab). c. R$_C$(H) / R$_C$(0) vs. H$\|$C for La$_2$Li$_x$Cu$_{1-x}$O$_4$ (T = 20 K; $x$ = 0.01 and $x$ = 0.04). d. R$_{ab}$(H) / R$_{ab}$(0) vs. H (T = 20 K; E$\|$ab and H$\|$ab) for both Li-doped samples.} \label{fig:s5}
\end{figure*}

\section*{S2. Electric polarization measurements: Pyrocurrent method}

To measure the electric polarization of our samples we employed the pyrocurrent method. The sample is placed between two metallic plates to form a capacitor. An electrometer and a voltage source are connected to form the electric circuit shown in Fig~\ref{fig:s2}a. The capacitor is placed into a cryostat, where we can accurately control the temperature. When the sample exceeds its ferroelectric transition temperature its electric polarization decreases abruptly to zero. This induces a current, which is called pyroelectric current or pyrocurrent.

Further to the pyrocurrent, resistive and capacitive parasitic electric signals are also included in the total measured current, which can be expressed by \cite{s3}

\begin{eqnarray}
i={V\over R} + C{\partial V\over \partial t} + V {\partial C\over \partial t} + A{\partial P\over \partial t}
\end{eqnarray}

Here V is the applied voltage, R and C are the resistance and the capacitance of the sample respectively, P is the electric polarization and A is the area of the capacitor. The first term corresponds to the resistance of the sample, and can be neglected as long as the sample is highly resistive or when the measurement is performed in V = 0 bias voltage. The second term relates to the charging current of the capacitor and is negligible for a zero voltage across the circuit. The third term is related to the current that could be caused when the dielectric permittivity of the sample changes considerably with temperature and can be ignored when a zero bias voltage is applied. The fourth term corresponds to the pyrocurrent component of the measured current. It follows that pyroelectric current measurements should be performed under zero bias voltage V = 0 to ensure that the first three parasitic terms do not contribute to the measured current, and thus the measured current will be purely due to the change of the electric polarization with respect to temperature. In this case we can write Eq (1) as follows:

\begin{eqnarray}
i=i_p=A{\partial P\over \partial t}=A{\partial P\over \partial T}{\partial T\over \partial t}
\end{eqnarray}

Hence, the pyrocurrent is proportional to the sweep rate of the sample's temperature. Thus we can measure pyrocurrent by keeping a constant sweep rate of the temperature (Fig~\ref{fig:s2}b). Integrating the pyrocurrent curves, with respect to time gives a quantitative measure of the charge induced due to the depolarization of the sample (Fig~\ref{fig:s2}c). Dividing by the area of the capacitor we obtain the electric polarization of the sample.

\begin{figure*}[]
\begin{minipage}{1.4\columnwidth}
\includegraphics[clip=true,width=0.99\columnwidth]{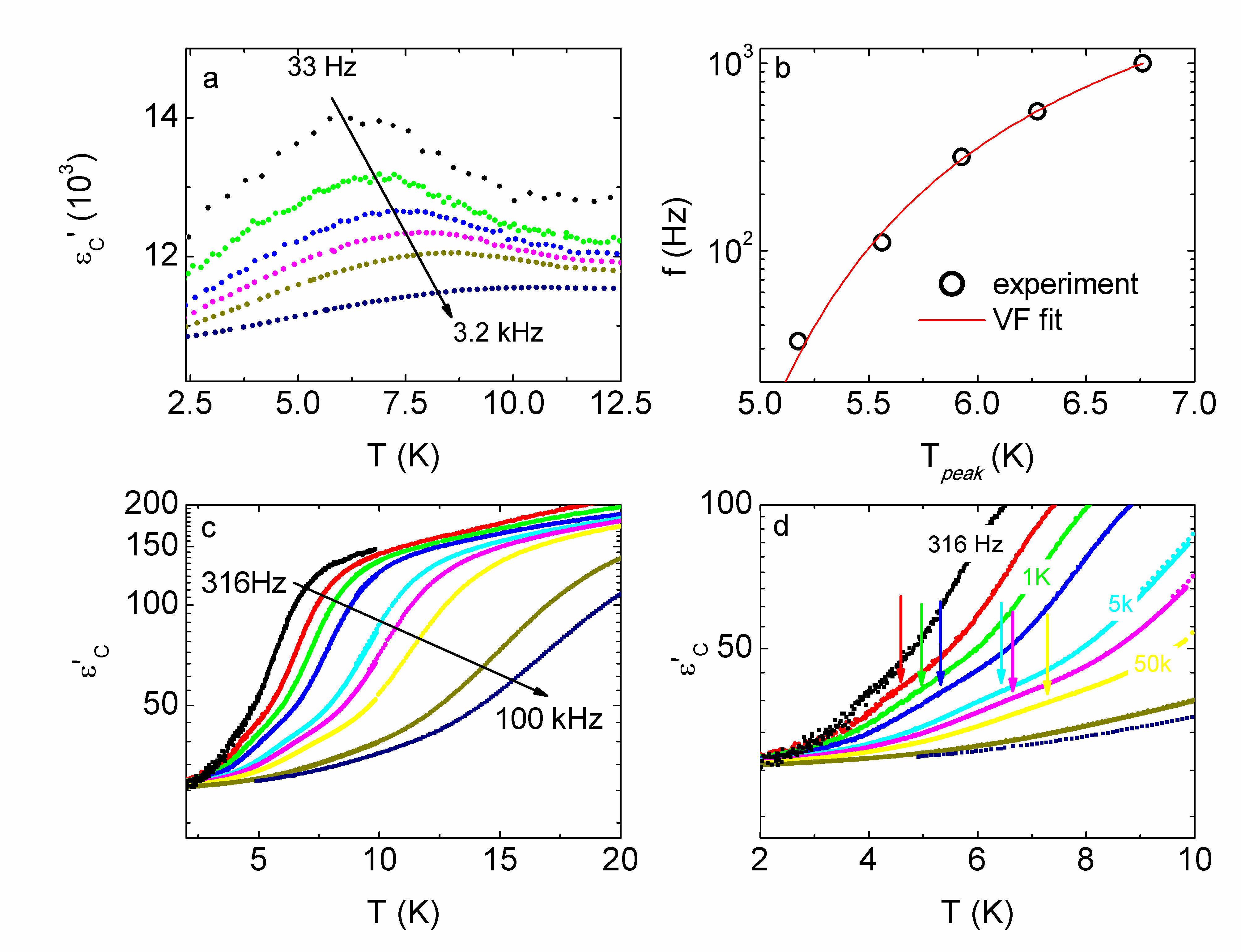}
\end{minipage}
\caption{(Color online) a. Out-of-plane dielectric permittivity $\epsilon'_C$  as a function of temperature for La$_2$Li$_{0.01}$Cu$_{0.99}$O$_4$. b. Frequency f as a function of peak temperature T$_{peak}$ (black open circles) fitted against the Vogel-Fulcher relation (red solid line).  c. Out of plane dielectric permittivity $\epsilon'_C$  for La$_2$Li$_{0.04}$Cu$_{0.96}$O$_4$. d. Additional changes in the slope of the permittivity are also observed and indicated by colored arrows.} \label{fig:s6}
\end{figure*}

\section*{S3. Measurement protocol}

To measure accurately the electric polarization of our samples we employ the following protocol. The measurement process begins with the sample well above its ferroelectric transition temperature (paraelectric state). An electric field is applied and the sample is cooled to 2 K. The electric field is removed to fulfill the conditions described in the former paragraph. The sample is then heated at 3 K/min while measuring time, temperature and current at a constant interval of 0.5 seconds. We continue to heat the sample well above its ferroelectric transition temperature. In the case of measuring the pyrocurrent in the presence of an applied magnetic field, we apply the magnetic field as soon as the sample reaches at 2 K and before removing the applied electric field (Zero Magnetic Field Cooling).

\section*{S4. Experimental data processing}

A typical measured current-vs.-time data set is shown in Fig~\ref{fig:s3}a. For T $>$ 7 K the current increases steeply with increasing temperature, indicating the conductive character of the sample. Although V = 0 during the measurement process, a small but unavoidable voltage arising from the voltage source is still applied to the circuit causing the observed resistive behavior - the level of this voltage depends on the current measurement range used in the current meter, and can be between $\mu$V and 200 mV. However, in the temperature range 2 K - 6 K there is a local current minimum (inset Fig~\ref{fig:s3}a) and its sign is reversed (Fig~\ref{fig:s3}b) upon reversing the electric field during the cooling process. Furthermore, it becomes sharper with increasing the temperature sweep rate suggesting its pyroelectric nature. Thus it is important to distinguish between the pyrocurrent and the resistive components of the current in order to accurately determine the electric polarization of the sample. To this point, we fit the experimental data above 8 K against a resistivity model.\cite{s4} The best fit is subtracted from the experimental data and the resulting curve is the pure pyroelectric component (Fig~\ref{fig:s3}c). Beyond 250 seconds there is considerable noise due to the data processing. Thus, the pyrocurrent curve is integrated with respect to time in the range 0 - 250 seconds, and the electric polarization with respect to time is acquired. We obtain P(T) by converting time to temperature.

\section*{S5. Dielectric permittivity of the lightly oxygen-doped L\lowercase{a}$_{2}$C\lowercase{u}O{$_{4+\lowercase{x}}$}}

We analyzed the dielectric permittivity data for the lightly oxygen-doped La$_2$CuO$_{4+x}$ (T$_N$ = 313 K) (Fig~\ref{fig:s4}).  The out-of-plane component of the permittivity $\epsilon''$c (Fig~\ref{fig:s4}a) exhibits a step-like decrement (at T $>$ 60 K) with decreasing temperature and shifts to higher temperature with increasing frequency indicating a common dipolar relaxation process. Furthermore, a dielectric peak is observed at T $\sim$ 20 K. This peak becomes smoother and shifts to higher temperature with increasing frequency, suggesting a charge relaxor-like similar to La$_{1.999}$Sr$_{0.001}$CuO$_{4+y}$. Figure~\ref{fig:s4}b shows the frequency f as a function of peak temperature T$_{peak}$ extracted from Fig~\ref{fig:s4}a. Application of the VF model gives a freezing temperature T$_{fr-c}$ = ( 8 $\pm$ 0.5 ) K. Similar analysis was performed for the in-plane component of the dielectric permittivity (Fig~\ref{fig:s4}c) and the calculated freezing temperature was found to be T$_{fr-ab}$ = ( 5.2 $\pm$ 0.4 ) K (Fig~\ref{fig:s4}d).

\section*{S6. Magnetoresistance measurements}

Figure~\ref{fig:s5} shows the magnetic field dependence of the normalized in-plane (Fig~\ref{fig:s5}b) and out-of-plane (Fig~\ref{fig:s5}a) resistivity of La$_{0.999}$Sr$_{0.001}$CuO$_{4+y}$ determined from measurements of electric impedance at T = 20~K. For H $\|$ c (Fig~\ref{fig:s5}a) we observe a first order phase transition and a corresponding hysteresis at H $\sim$ 6 T due to the metamagnetic transition associated with DM interactions. In La$_2$CuO$_{4+x}$, the crystal anisotropy and the DM interaction fix the easy axis for the spins to the longer of the two in-plane orthorhombic directions (the b-axis). The direction of the (WF) moments $\vL$ induced by the DM interaction is fixed by the cross product $\vL=\vD\times \vn_0$ between the DM vector $\vD$ (oriented along the shorter of the two in-plane orthorhombic directions - the a-axis) and the AF order parameter $\vn_0$ (pointing along the b-axis) so that $\vL$ is oriented along the c-axis, perpendicular to the CuO$_2$ planes of the crystal structure. A sufficiently large magnetic field applied along the c-axis can overcome the inter-plane AF coupling and induce a discontinuous spin-flop reorientation, causing the so-called WF (first order) phase transition. The critical field is reduced at high temperatures following the decrease in $\vL$, due to thermal fluctuations in $\vn_0$(T) [45]. For H$\perp$ab (Fig~\ref{fig:s5}b), the magnetoresistance varies smoothly because the weak ferromagnetic moments induce a continuous rotation of $\vn_0$ in the bc-plane. Similar results are obtained for the Li doped samples, as shown in Fig~\ref{fig:s5}c and Fig~\ref{fig:s5}d.

\begin{table*}[]
\begin{tabular}{|m{3 cm}|m{2.3 cm}|m{3.7 cm}|m{1 cm}|m{1.2 cm}|}
  \hline
  % after \\: \hline or \cline{col1-col2} \cline{col3-col4} ...
  & Estimated Carrier \newline Concentration \newline (cm$^{-3}$) & Dielectric permittivity \newline (@ 1kHz) & T$_{FE}$ \newline (K) & P$_c$ \newline(@ 2 K)\\  \hline
   La$_2$CuO$_{4+x}$ \newline (T$_N$=320 K)& 10$^{17}$ & ~2000 \newline (50 K - out of plane) & 4.5 & 33 \\ \cline{3-5}
    PRB (2012)\newline &   & ~3000 \newline (50 K - in plane) & 4.5 & 26 \\ \hline
  La$_2$CuO$_{4+x}$ \newline (T$_N$=313K)& ~10$^{18}$ & ~1200 \newline (100 K - out of plane) & 5 & 30 \\ \cline{3-5}
   Present study \newline &   & ~1300 \newline (50 K - in plane) & 3.5 & 27 \\ \hline
  La$_{1.999}$Sr$_{0.001}$CuO$_4$ \newline (T$_N$= 312 K) & ~10$^{18}$ & ~ 70 K \newline (100 K - out of plane) & 6.5 & 36 \\ \cline{3-5}
  Present study \newline &  & ~ 290 \newline (50 K - in plane) & 4 & 18 \\
  \hline
\end{tabular}
 \caption{Summarized values regarding the dielectric permittivity and electric polarization for both oxygen and Sr doped La$_2$CuO$_4$ samples. Corresponding values from our previous investigation on oxygen excess doped La$_2$CuO$_{4+x}$\cite{s6} are also included. }
\end{table*}

\section*{S7. Dielectric permittivity of L\lowercase{a}L\lowercase{i}$_{1-x}$C\lowercase{u}O$_{4}$}

Similar to La$_{0.999}$Sr$_{0.001}$CuO$_{4+y}$, we also analyzed the dielectric permittivity for La$_2$Li$_{0.01}$Cu$_{0.99}$O$_4$ (Fig~\ref{fig:s6}a). Dielectric peaks are observed in the low temperature regime, which are suppressed and move to higher temperature with increasing frequency. The intrinsic origin of the peaks was confirmed as in the case of La$_{1.999}$Sr$_{0.001}$CuO$_{4}$. VF fits give T$_{fr-c}$ = ( 5.0 $\pm$ 0.3 ) K (Fig~\ref{fig:s6}b). Correspondingly, measurements of the electric polarization discussed in the main text give a FE transition temperature T$_{FE}$ = 5 K.

For La$_2$Li$_{0.04}$Cu$_{0.96}$O$_4$, we observe step-like changes in the out-of-plane dielectric permittivity, which are suppressed and move to higher temperature with increasing frequency (Fig~\ref{fig:s6}c). Such features correspond to a normal dielectric relaxation process.  On the other hand, there are no observable peaks corresponding to a relaxor-FE behavior. For T $>$ 316 Hz we observe an additional small change in slope below the step-like transition (Fig~\ref{fig:s6}d). This feature shifts to higher temperature with increasing frequency and vanishes for f $>$ 10 kHz. Repeating the experiments several times with renewed contacts resulted to curves with similar features, indicating their intrinsic character. We cannot extract a corresponding f vs.T$_{peak}$ diagram since it is not possible to define the T$_{peak}$ hence, we cannot apply the VF model to determine a freezing temperature.\\


\begin{thebibliography}{99}
\bibitem{30} Vogel-Fulcher analysis can be applied on either the real or the imaginary part of the dielectric permittivity giving similar charge freezing temperatures. Since the dielectric peaks of the real part of the out-of-plane dielectric permittivity are not observable, we analyzed the imaginary part of the dielectric data where the corresponding peaks can be identified.
\end{thebibliography}
\begin{thebibliography}{1}
\bibitem{s4} The variable range hopping mechanism of conduction $\sigma = \sigma_0\exp(T_o/T)^{-a}$ is the most suitable mechanism to explain the hopping conduction in insulating antiferromagnets.
\end{thebibliography}
\end{document}